\def \msyr{$M_{\sun}~{\rm yr}^{-1}$}
\def \cm{~\rm{cm}}
\def \s{~\rm{s}}
\def \km{~\rm{km}}
\def \kms{~\rm{km}~{\rm s}$^{-1}$}

\def \g{~\rm{g}}

\def \erg{~\rm{erg}}

\def \yr{~\rm{yr}}

\documentclass[onecolumn]{aa}
\usepackage{graphics}

\begin{document}


\title{Launching jets from the boundary layer of\\
    accretion disks in young stellar objects\thanks{Research supported by
    the Israel Science Foundation,
    the Technion Fund for the Promotion of Research and the Helen \& Robert Asher Fund} }

\author{Noam Soker\inst{1,2}
      \and Oded Regev\inst{2,3}}

\offprints{Noam Soker, \email{soker@physics.technion.ac.il}}

\institute{Department of Physics, Oranim,
  36006 Tivon, Israel \and
Department of Physics, Technion-Israel Institute of
Technology,
  32000 Haifa, Israel \and
Department of Astrophysics, American Museum of Natural History,
New York, NY 10024, USA }

\date{Received ---- / Accepted ----}

\titlerunning{Jets from boundary layers}

\abstract{
We reexamine a previously proposed model for thermal pressure
acceleration of collimated outflows in young stellar objects (YSO).
We are motivated by new results from recent X-ray observations
of YSO. These show that there is essentially no difference between
the properties of X-ray emission from YSO with and without outflows,
imposing quite severe constraints on models based on magnetic launching
of jets.
In our scenario the magnetic fields are weak, and serve only to recollimate
the outflow at large distances from the source.
We perform time scale estimates and an analytical calculation of
the acceleration of gas originating in the boundary layer (BL) of an accretion disk.
By applying global energy conservation considerations
we find that the mass escaping the system is compatible with observations.
A crucial ingredient of the proposed model is that the accreted
material is strongly shocked, and then cools down on a time scale
longer than its ejection time from the disk.
By using appropriate properties of accretion disks boundary layers
around YSO, we show that the conditions for the proposed scenario to
work are reasonably met. We find also that the thermal acceleration mechanism
works only when the accretion rate in YSO accretion disk is large enough and
the $\alpha$ parameter of the disk small enough - otherwise the cooling time
is too short and significant ejection does not take place.
This result appears to be compatible with observations as well
as theoretical considerations.}
\maketitle



\section {Introduction}
\subsection {General}

Energetic outflows (in particular collimated jets) and  accretion flows (in particular
accretion disks) seem to be among the most prominent signatures of star formation.
The discovery of large molecular outflows dates back to the CO observations of
Snell et al. (1980) while that of jets to observations of optical lines
by Mundt \& Fried (1983) and of the radio continuum by Bieging et al. (1984).
Since then, these important phenomena
have been extensively studied and classified (for recent reviews on the
different types and aspects of these outflows see Eisl\"offel et al. 2000,
Richer et al. 2000, Reipurth \& Bally 2001) and it is now rather generally
accepted that the fast and collimated jets actually drive the slower, more massive,
large molecular outflows.

 The presence of disks in young
stellar objects (YSO) has at first been inferred indirectly, to account for
excess emission in the IR, UV and millimeter bands (see e.g. Bertout 1989, Hartigan et al.
1995). Direct evidence for the presence of disks has recently become available as well,
using millimeter and submillimeter interferometric mappings (see Wilner \& Lay 2000),
adaptive optics in the near infrared and the HST in the optical (see McCaughrean {\em
et} al. 2000).

It became clear almost at the outset that "conventional" stellar winds, emanating
from the protostar, alone can not
power the energetic outflows (DeCampli 1981, K\"onigl 1986). Moreover,
strong phenomenological correlations between observational signatures of outflows
and of accretion (and disks in particular) have been found (Cabrit
et al. 1990). By now, these correlations, together with subsequent theoretical work,
have developed into what is widely known as the accretion-ejection paradigm, a theoretical
framework based on the idea that energetic
outflows are actually physically linked to accretion disks and that it is accretion
that actually powers these outflows and notably the fast collimated jets. A very recent
comprehensive phenomenological summary on the accretion-ejection structures in YSO and the
observational constraints on possible theoretical models can bee found in Cabrit (2002).

The above discussion actually concerns solar mass type, low luminosity
($L_{\rm bol} \la 10^3 L_{\sun}$) YSO, but it appears that this picture
is also relevant to high mass, luminous ($L_{\rm bol} \ga 10^3 L_{\sun}$)
YSO (see K\"onigl 1999).
Moreover, the accretion-ejection paradigm seems to apply to other
occurrences of astrophysical jets as well, but in this paper we
shall limit ourselves to dealing solely with the low luminosity
YSO case and hence will neither review or discuss other cases nor propose
any unified schemes.

On the theoretical side, a rather wide consensus has by now been
established assigning to magnetic stresses alone the role of mediating
between the above mentioned inflow and outflow processes
(however, see Hujeirat, Camenzind \& Livio 2003, who consider thermal
as well as magnetic effects for launching jets in quasars).
A link of this kind, postulated to occur on large scale,
is thought to contain in it also the physical mechanism for the
ejection of matter from the accretion disk and for its collimation.
This view is usually based on global conservation considerations
using observational estimates of the outflow's momentum discharge $\dot P_{\rm o} \equiv
\dot M_{\rm o} v_{\rm o}$ and its
kinetic luminosity $L_{\rm kin} \equiv (1/2)\dot M_{\rm o} v_{\rm o}^2$,
where $\dot M_{\rm o}$ indicates the mass loss in the outflow and
$v_{\rm o}$ its velocity. In particular, it has been found that the
momentum discharge in the outflows exceeds by a factor of $100-1000$
the radiation pressure thrust of
the central source ($L_{\rm bol}/c$) (e.g. Lada 1985) and therefore
the flow can not be driven by radiation.
Moreover, the ratio of $L_{\rm kin}$ to $L_{\rm bol}$
(which is largely due to accretion) is up to $\sim 0.1$, indicating a rather
efficient conversion of the gravitational energy, liberated from the accreting material,
into the kinetic energy of the outflow.
Such high ejection efficiency is thought to be most naturally understood if the outflows are
driven magnetically.

The strong collimation of some of the jets is also naturally interpreted as resulting from the
operation of dynamically significant magnetic fields. For example, high resolution HST
observations of a typical disk/jet protostellar system HH30 (Burrows et al. 1996)
indicate that it is well collimated already
very close to the source and moreover appears to recollimate
further out. For Ferreira (2002) this observation was enough to conclude
that the jet must be self collimated and this, in turn, indicates that the only viable physical
jet model must rely on the action of a large scale magnetic field, driving the jet and being carried
by it.
The collimation issue is, however, still quite controversial
and while magnetic collimation is certainly plausible, its exact nature is probably
quite involved and still not fully understood (see the recent works of Bogovalov \& Tsinganos 2001
and Li 2002).

\subsection {Magnetic ejection and X-ray emission}

The currently accepted scenarios, devised to explain the accretion-ejection
paradigm in the case of bipolar outflows in low mass YSO, is therefore
based on the operation of large scale magnetic fields driving the flow
from the disk; either via the "centrifugal wind" mechanism, first proposed
by Blandford \& Payne (1982), or from a narrow region in the magnetopause
of the stellar field via an "X-wind mechanism" introduced by Shu et al.
(1988, 1991) and in a somewhat different setting, by Ferreira \& Pelletier
(1993, 1995). The details of the various magnetic accretion-ejection
theoretical models are certainly beyond the scope of
this contribution and we refer the reader to the recent
reviews by K\"onigl and Pudritz (2000), Shu et al.
(2000) and Ferreira (2002) for detailed discussions of this issue.
It should, however, be pointed out that the origin of the large scale
magnetic fields and the manner that open filed lines of sufficiently strong
magnitude persist (in the centrifugal wind models), or the manner by which
a stellar field interacts with the disk, allowing inflow and
at the same time driving an outflow (in the X-wind models) are still
open key issues of the theory.

In any case, in a model based on magnetic driving it is to be expected,
by definition, that the magnetic energy density (pressure) plays
dynamically important role over sufficiently large regions. A naive
expectation in such a situation, based partially on the behavior of
the solar wind, is that the strong magnetic fields, which are probably
naturally prone to various instabilities, may undergo resistive reconnection
and lead to copious production of X-rays. In the sun, where the wind
properties are determined by the magnetic activity, the kinetic luminosity
of the solar wind is of the order of the average solar X-ray luminosity
(although most of the dissipated energy in the magnetic activity goes
 to heat the plasma).

In a very recent report on extensive {\em Chandra} X-ray observations of YSO, Getman et al. (2002)
find that the average X-ray luminosity of seven out of eight YSOs with outflows is
$L_{\rm x} \la 3 \times 10^{29} \erg \s^{-1}$ and therefore significantly below the typical kinetic
luminosity in YSO jets. The latter can be parameterized as follows
\begin{equation}
L_{\rm kin} = 2 \times 10^{31}
\left( \frac{\dot M_j}{10^{-9} M_{\sun} {\rm yr}^{-1}} \right)
\left( \frac{v_j}{250 \km \s^{-1}} \right)^2
\erg \s^{-1} ,
\label{ek11}
\end{equation}
and therefore if we adopt estimates of jet velocities in the range $ 150 \la v_{rm j} \la 400$ \kms and
of mass loss $10^{-10} \la \dot M_{\rm j} \la 10^{-8}$ \msyr(see e.g. K\"onigl \& Pudritz 2000), we get
\begin{equation}
50 \la \frac{L_{\rm kin}}{L_{\rm x}} \la 500.
\label{ratio}
\end{equation}
More importantly, Getman et al. (2002) find that the presence
or absence of an outflow does not appear to produce any difference in
the X-ray properties of YSO.
We find this lack of observational correlation rather surprising if
the outflows are indeed dynamically driven by magnetic fields.

The fact that X-ray luminosities of TTauri stars are low, i.e. in
the range of $10^{-3}-10^{-4}$ of {$L_{\rm bol}$), the bolometric luminosity,
(for both CTTS and
WTTS, that is, those who are though to still accrete from circumstellar
matter and those which are "naked") has already been discovered in the
early 80-s by the {\em Einstein} observations (Feigelson \& DeCampli 1981).
As $L_{\rm kin}$ of the outflow is of the order of $0.1-0.01$ of
$L_{\rm bol}$ (see above) the new {\em Chandra} findings, which give the
luminosity ratio (eq. \ref{ratio}),
are consistent with the earlier {\em Einsten} observations.
The similarity of various correlations related to this X-ray activity in
TTauri stars and in ordinary solar type stars has lead to the assumption
that this X-ray emission can be attributed to some kind of solar-type
coronal heating.
However, as was first pointed out by DeCampli (1981), the low X-ray
luminosity probably rules out a similar (to the solar case)
mechanism for driving a TTauri wind.
A related observation is that although TTauri stars have enhanced
(over comparable normal late-type stars) optical and UV emission lines,
their X-ray fluxes are quite similar.
This is referred to as {\em X-ray deficit}. See Bertout (1989)
for a comprehensive account on these matters and references.

It seems to us that this kind of ratio value would not be disturbing
in sources without outflows, or if in a magnetically driven outflow the
large scale magnetic field was in some, more or less, steady state.
However, such an assumption must impose severe (and probably even
prohibitive) constraints on the magnetic configuration
in the accretion-ejection model. This is so because,
after all, the outflow energy source is in the accreted matter from
the disk, which releases its gravitational energy. Since the
magnetic field is assumed to be linking the disk and the outflow,
namely the matter to be accreted and that which is being ejected,
it is reasonable to conclude that a significant fraction of the magnetic
field energy is somehow dissipated in the ejection process, be it
in the disk, on the YSO surface and/or in the outflow. Why then is the
X-ray luminosity of YSO with outflows so low and, in particular,
why there are no differences between the X-ray luminosities of YSO with
and without outflows?

 It is perhaps instructive to elaborate more on the reason why we
find the above mentioned X-ray results to be a sufficient motivation
for proposing a new scenario.
The accretion disk consists of a strongly sheared flow and, in addition,
matter is accreted through it and therefore there is also a radial inflow.
In the different accretion-ejection models (see the references above)
the origin of the magnetic field is assumed to be the central star, or
the primordial cloud or both, but in all these models some kind of
a dissipative process is invoked in order to sustain a quasi-steady magnetic
field configuration (as is also necessary for magneto-centrifugal ejection).
Effective field diffusivity to allow for the stellar field slippage through
the disk (attributed perhaps to turbulence) has already been assumed in models
based on the Ghosh \& Lamb (1978) picture. Without invoking such processes
the field becomes extremely twisted and pinched towards the star as well -
a configuration leading to strong reconnection. These models do not treat
mass ejection and concentrate in the star-disk coupling and accretion problem.
K\"onigl (1991) used the Ghosh \& Lamb model and applied it to protostellar
accretion (a detailed account on the physics of magnetized accretion disks
can be found in the book by Campbell 1997).

In the accretion-ejection models, and in particular in those based on the
magneto-centrifugal ejection (Blandford \& Payne 1982), effective plasma
conductivity (ambipolar diffusion or effective Ohmic diffusivity resulting
from turbulence) is assumed as well (see K\"onigl 2000 and references therein).
Such an effective conductivity must give rise to an effective dissipation,
comprising of local turbulent reconnection events which can be effectively
cast into a ``macroscopic'' Ohmic dissipation parameter
(analogously to what is done in the $\alpha$
prescription for effective disk viscosity, also leading to dissipation).
Thus energies of the order of the magnetic field energy density are
dissipated by the flow, not in pure large scale ``reconnection'' events,
but rather by means of ``effective'' field diffusion and dissipation.
It ammounts to the same, not in prominent flares, but in heating
which should give rise to X-rays.  These processes must operate in the disk.
Outside the disk, matter may be flowing along field lines and
reconnection or effective turbulent dissipation do not have to occur.
In a recent numerical study, K\"uker, Henning \& R\"udiger (2003) have
included such eddy magnetic diffusivity in their calculations but still
concluded that an ultimate steady field configuration is extremely
unlikely.
In any case, all what we claim is that a dynamically active magnetic field,
which extends from the disk and outward, should suffer from significant
dissipation, which we expect to result in heating and a relatively strong
X-ray emission, contrary to observations.

In reviewing different astrophysical objects possessing or lacking jets
and/or accretion disks, Livio (1999, 2000) proposed the conjecture that
powerful jets are produced by systems in which on top of an accretion
disk threaded by a vertical field, there exists an additional source
of energy/wind (possibly associated with the central object).
We find this conjecture very reasonable in view of the
X-ray observations of YSO mentioned above and our interpretation of
their possible significance. It is important
to stress that we do not want to claim that magnetic fields are not important
in the physics of outflows from YSO. Indeed, magnetic fields of significant
strength have been indirectly and even directly observed in the vicinity
of YSO and further out, in the outflows (e.g. Crutcher et al. 1993,
Ray et al. 1997). In particular, values of the order
of $1 {\rm kG}$ have been reported on the stellar surfaces, most probably
localized in stellar spots (e.g. Basri et al. 1992, Guenther et al.
1999, Johns-Krull et al. 1999).
Their role in channeling and
collimating the outflow should be quite obvious from these observations (for
recent summaries on the various results see the reviews by K\"onigl 1999,
K\"onigl \& Pudritz 2000, Cabrit 2002).

\subsection{Ejection by thermal pressure}

In this work we shall try to examine
a non-magnetic (and explicitly thermal) source for the {\em  ejection} of
material from very close to the surface of the central object,
or the inner disk edge.

The natural candidate for the location of such a source is the accretion disk boundary
layer (BL), a narrow region in which the Keplerian accretion flow has to
rapidly adjust to the presence of a physical inner boundary
of the disk (be it the surface of the
star or some constraint imposed by the inner, i.e. stellar, magnetic structures).
As such, the BL is a region in which the accretion flow is highly sheared
and is subject to increased dissipation. Semi-analytical asymptotic, as well as numerical,
models of steady, one-dimensional viscous accretion disk BLs, have been introduced
by Regev 1983, Popham et al 1993, Regev \& Bertout 1995 and others (see Popham 1997
for a fairly recent critical summary on the questions concerning these models). Full,
time dependent, two-dimensional numerical calculations have also been attempted, recently
achieving a run-time of about 1000 orbital periods, still significantly too short for
obtaining a final steady or quasi-steady configuration (Kley \& Lin 1996).
These calculations were non-magnetic and did not model any mass loss from the BL, but it
is interesting to note that Bertout \& Regev (1992) could find steady solutions of
optically thick, cool BLs only if significant mass loss from the BL was
incorporated.

In an earlier work, Pringle \& Savonije (1979) examined the possibility
of explaining the X-ray emission of dwarf-novae by invoking shock dissipation
in the BL. The above mentioned viscous models did not allow for shocks but it is
certainly reasonable that the strongly sheared turbulent accretion flow in the BL
is likely give rise to shocks, possibly strong ones.

Pringle (1989) himself also proposed that the most likely origin for
collimated winds in YSOs is the BL of accretion disks.
However in his model it is the strong toroidal magnetic field
generated in the layer that is being responsible for the ejection. We expect
that also in this model the twisted magnetic field is likely to produce significant
X-ray emission, above the values observed in YSO or at least significantly
more that in YSO without mass loss.

We are thus led to the examination of a possibility
to drive outflows from the vicinity of the central object, specifically the disk
BL, by thermal pressure.
This obviously depends on the relation between the heating and radiative
cooling times of the relevant physical processes in the region in question.
It has generally been accepted that the relevant cooling times are
sufficiently short, so that a quick and significant pressure buildup
(as a result of dissipation) and acceleration of matter to escape
velocities is highly unlikely.
Indeed, estimates of {\em viscous} heating vs radiative cooling in
various conditions and cases enable vertical hydrostatic equilibrium
of the disk and of the BL as well (see the above cited works).
However, if the flow in the BL is turbulent
and blobby, we may expect that  spatiotemporally localized accretion
(strong) shocks (SPLASH) may occur. If such regions of shocked material
are not able to cool sufficiently fast we may reasonably expect violent
expansion of localized regions and ejection of matter.
The idea that energy liberated in a BL shock may thermally drive bipolar
winds was proposed almost 20 years ago by Torbett (1984).
Later on, Torbett \& Gilden (1992; hereafter TG) performed numerical calculations,
solving the hydrodynamical equations in one dimension
(vertical) with the BL being arrested in a single strong shock. They found that without
radiative losses there is mass ejection but when the latter are taken into account matter
can be accelerated up to only approximately 0.25 of the escape velocity.

In this paper we shall reexamine this model and suitably modify it in order to
assess the feasibility of a thermal pressure ejection mechanism of matter
from accreting YSO. In doing so we shall try to place constraints on
possible models of this kind.

\section {Mass ejection from SPLASHes in the boundary layer}

Prompted by the considerations given in the Introduction,
we examine here the possibility of matter ejection from the
accretion disk BL in YSO as a result of violent dissipation events,
which give rise to rapid expansion.
To achieve such circumstances we postulate
that spatiotemporally localized accretion shocks (SPLASH) are formed
in the BL and that these shocks are sufficiently strong. Spatiotemporal
localization means that the amount of matter shocked in an event is
relatively small (a fraction of the order $b/R$,
where $b$ is the boundary layer's width and $R$ is the stellar radius,
of the matter in the BL) and that
the dissipation event is of short duration (of the order of the dynamical
time in the BL, i.e. hours to days). Here and henceforth we shall use
the term SPLASH to denote both the strong shock itself and the BL region
shocked by it.

If SPLASHes of this kind give rise to matter ejection in
the vertical direction and if they repeat themselves with a
sufficiently high frequency (consecutive events can originate from
different parts of the BL) they can form a large jet segment, and
when they are observed they can not be resolved.
The observational manifestation of such a process may be in small
variations in the luminosity of the central source on time scales
of hours to days.
Such variations, thought to result from obscuration and/or changes
in mass accretion rates, have indeed been observed in different YSOs
(e.g. Skurtskie et al. 1996, de Winter \& van den Ancker 1997).

Let us consider a geometrically thin accretion disk BL assuming, for the
sake of convenience, that it is isothermal in the vertical direction.
Viscous steady accretion disks are thin when they are relatively cold,
that is, when the sound speed $c_s$ is much smaller than the Keplerian
velocity $v_{\rm K}$ at the point in question. Standard disk theory
(see Pringle 1981) then gives that the disk's vertical extension
$h$ at a given radial location $r$ is estimated to be $h = \epsilon r$,
where $\epsilon \equiv c_s/v_{\rm K} \ll 1$ at $r$.

The BL's radial extension, $b$, has been estimated in a number of
theoretical models of viscous and steady BLs to be between $\epsilon^2 R$
and $\epsilon R$, depending on the nature of the BL (e.g. Pringle \& Savonije 1979,
Regev 1983, Popham et al. 1993, Regev \& Bertout 1995).
In the present context we expect the formation of a rather hot and broad
BL (see below) and therefore the estimate $b= \epsilon R = H$,
where $H$ is the disk height at the BL, i.e.  $H=h(R)$ (see Regev \& Hougerat 1988,
Regev \& Bertout 1995, Popham 1997), seems appropriate.

In steady state the vertical structure of the disk (and of the BL as well)
is governed by the hydrostatic equation
\begin{equation}
 \frac {1}{\rho} \frac {d P}{dz}   = -\frac {z}{r^2} v_K^2,
\label{hydro1}
\end{equation}
where $P$ is the pressure, $\rho$ the density and $z$ denotes the vertical
coordinate.
The density profile can be derived by solving this equation when a pressure
density relation is assumed (or obtained from additional equations).
As mentioned above we assume that
the BL is isothermal in the vertical direction, and therefore $P = c_s^2 \rho$
with $c_s$ being $z$ independent.
The isothermal assumption sufficient for our purposes, because we do not aim at
detailed modeling. In any case, the deviations from isothermality in vertical disk
structure are not too large in the bulk of the disk.
Thus,
\begin{equation}
\rho=\rho_0 \exp \left( - \frac {v_K^2}{c_s^2} \frac {z^2}{2 r^2} \right).
\label{dens1}
\end{equation}

In steady viscous disk models it is assumed that the effects of turbulence
in the disk and the BL can be effectively parameterized by some sort of a
turbulent viscosity which scales with the disk's height $h$ and the sound
speed $c_s$. Thus the calculated structure is actually a suitably defined
average. Because the BL is more strongly sheared than the disk further out,
it is reasonable that it is a site of vigorous activity of matter "blobs",
occasionally colliding with each other. Such collisions are bound to
create shocks which cause the shocked regions to expand in all directions.
As long as these shocked regions are sufficiently small (with respect to
the size of the system, i.e. the BL) the approach of assuming an overall
steady average state remains viable. However, if
the shocked regions continue to expand out into the path of yet more
circulating blobs, stronger shocks may be created.
This kind of scenario was proposed by Pringle \& Savonije (1979)
to explain the emission of X-rays out of disk BLs in dwarf novae.
In our case the interaction of the orbiting BL blobs with magnetic
structures protruding from the star may also play a role in forming
strong shocks of this kind.
The feasibility of the formation of such spatiotemporally
localized (but not too small!) accretion shocks, which we call SPLASHes,
obviously depends on the ratio between the relevant cooling time
of {\em individual blobs}, $t_{\rm cool}$,
and the adiabatic expansion time $t_{\rm ad} = \ell/c_s$, where $\ell$
is the size of the expanding blob.
The condition
\begin{equation}
t_{\rm cool} \ga t_{\rm ad}
\label{splashform}
\end{equation}
is necessary for the formation of SPLASHes themselves
(compare with Pringle \& Savonije, 1979) and also, as we shall see later,
the ability of a SPLASH to accelerate non-negligible amounts of matter
beyond escape velocity depends on a number additional conditions.
We defer the estimation of these important time scales to the
next section, where we shall discuss them in detail.
It is important to distinguish between weak shocks resulting from
collisions between relatively small blobs, which orbit the star
and whose relative velocities are only slightly supersonic,
and the strong shock which we invoke to be the result of the merger
of several hundred expanding blobs to form a large disturbance in the disk
(see next section).
As individual blobs collide, the postshock temperature is much too low
to expell matter from the disk.
However, the collisions of more and more {\em expanding}
blobs, creates a fast growing disturbance until the disturbed region is large
enough so that it must be reasonably described by the physics of a strong
shock, where most of the disk's kinetic energy is converted to thermal energy.
The buildup of the large disturbance region from small blobs is such that
not all accreting disk material passes through theses large disturbances
(SPLASHes), and not all the resulting thermal energy is converted into
kinetic energy of the ejecting matter. However, as we calculate below,
such a process can work and effectively accelerate only a (rather small)
fraction of the matter to above escape velocity

Assume that a SPLASH occurs in the BL of an accretion disk around a YSO, that is,
a region of size $ \sim b = H$ is strongly shocked and therefore
its temperature rises by a large factor, $\eta \gg 1$ (see Torbett 1984)
in a very short time.
Assuming an isothermal disk both before and after the occurence of this SPLASH,
we find that the vertical acceleration as a function of $z$ after the
SPLASH is
\begin{equation}
a_z=
- \frac {1}{\rho} \frac {d P}{dz}
-\frac {z}{r^2} v_k^2 = \frac {z}{r^2} v_k^2 (\eta -1),
\label{accel1}
\end{equation}
where we have used the fact that the pressure gradient term after the SPLASH,
i.e. in equation (\ref{accel1}), is actually $\eta$ times the value of
that term before the SPLASH, i.e., in equation (\ref{hydro1}).
Assume now that the shocked matter expands a distance $\sim z$ before substantial
cooling occurs (see below) and therefore the acceleration is approximately constant.
During this expansion the material reaches a velocity of $ v_z \sim (2 z a_z)^{1/2}$.
We parameterize this as
\begin{equation}
v_z = \beta (2 z a_z)^{1/2} =
\frac {z}{r} v_{\rm esc} (\eta -1)^{1/2} \beta,
\label{vz1}
\end{equation}
with $\beta \sim 1$ being $z$-independent, and
$v_{\rm esc} = 2^{1/2} v_K$ is the escape velocity from a distance
$r$  (assuming $z \ll r$).
A constant value of $\beta$ with $z$ is justified by the
self similar expansion found by TG.
Torbett (1984) parameterized the velocity in a different way,
including a parameter allowing for radial expansion as well
as vertical one, and the adiabatic index to account for
adiabatic cooling.
We prefer the above parameterization because we differ from Torbett
(1984) in using a global energy consideration (see below) which we find more
appropriate.

As the result of a SPLASH the kinetic energy of the Keplerian motion
of the shocked material in the BL is converted to thermal energy and
then back to kinetic energy via the expansion described above.
The residual thermal energy of the gas in the BL after the expansion
and that before the SPLASH are clearly very small compared to the kinetic
energy. In addition, because the acceleration is assumed to act only
along a short distance (Torbett 1984), the difference between the gravitational
energy of the matter before the SPLASH and after the expansion is negligible
as well. Thus global energy conservation implies that the kinetic
energy of the initially Keplerian-rotating material is
fully (approximately) converted to kinetic energy of the wind if there
are no radiative losses. These clearly should be take into account, however.
They can be conveniently parameterized if we assume that globally
a fraction $\Gamma < 1$ of the initial kinetic energy is converted to
the kinetic energy of the escaping material and the remainder fraction
of $1-\Gamma$ is lost to radiation.

The initial Keplerian kinetic energy per unit disk area on the disk
surface is
\begin{equation}
E_{ki}=\frac {1}{2} v_K^2 \int_{-\infty}^{\infty} \rho dz
\label{ek0}
\end{equation}
while the kinetic energy of such a unit area of the expanding material is
\begin{equation}
E_{kf}=\frac {1}{2} \int_{-\infty}^{\infty} v_z^2 \rho dz,
\label{eke}
\end{equation}

Taking $v_z$ from equation (\ref{vz1}) and $\rho$ from equation
(\ref{dens1}), the relation $E_{kf}=\Gamma E_{ki}$, which is actually
the definition of $\Gamma$, gives
\begin{equation}
4 \beta^2 (\eta-1) c_s^2 \int_{-\infty}^{\infty} x^2 e^{-x^2} dx =
\Gamma v_K^2 \int_{-\infty}^{\infty} e^{-x^2} dx ,
\label{eta1}
\end{equation}
where we define
\begin{equation}
x\equiv \frac{z}{r} \frac{v_K}{2^{1/2} c_s} .
\label{xdef1}
\end{equation}
This can be integrated analytically to yield
\begin{equation}
\beta (\eta-1)^{1/2}  = \Gamma^{1/2} \frac {v_K}{2^{1/2} c_s}.
\label{beta1}
\end{equation}
Substituting this result back in equation (\ref{vz1}) gives
\begin{equation}
v_z = \Gamma^{1/2} \frac {z}{r} \frac {v_K^2}{c_s} =
\Gamma^{1/2} x v_{\rm esc}.
\label{vz2}
\end{equation}

The condition for the material to escape is $v_z \ge v_{\rm esc}$,
which by the last equation is equivalent to
\begin{equation}
x= \frac {z}{r} \frac {v_K}{2^{1/2} c_s} \ge \Gamma^{-1/2}
\label{zm1}
\end{equation}
The mass in the accretion disk (actually the BL) which resides
close to the disk's plane, so that the last inequality does not hold for it,
will not reach escape velocity.
A fraction of the mass, however, reaches very high speeds,
and escapes the BL, mainly along in the $z$ direction.
The mass per unit area reaching the escape velocity is given by
\begin{equation}
\Sigma_{\rm esc} = 2 \int_{z_m}^{\infty} \rho dz,
\label{sigma1}
\end{equation}
where $z_m$ is the value of $z$ obeying the equality sign
in equation (\ref{zm1}), and
the factor 2 comes from counting both sides of the disk.
Changing the variable to $x$, we get the fraction of the disk material
reaching escape velocity to be
\begin{equation}
f_m= \left( \int_{\Gamma^{-1/2}}^{\infty} e^{-x^2} dx \right)
 \left( \int_0^{\infty} e^{-x^2} dx \right)^{-1}.
\label{fm1}
\end{equation}

The fraction of the total energy that is carried
by the escaping mass is derived from equation (\ref{eke}) as
\begin{equation}
f_E= \Gamma \left( \int_{\Gamma^{-1/2}}^{\infty} x^2 e^{-x^2} dx \right)
       \left( \int_0^{\infty} x^2  e^{-x^2} dx \right)^{-1} .
\label{fe1}
\end{equation}

The momentum per unit mass carried by the wind (on one side of the disk)
is given by
\begin{equation}
v_a= \frac {\int v_z \rho dz }
       {\int \rho dz } =
\Gamma^{1/2} v_{\rm esc}
\left( \int_{\Gamma^{-1/2}}^{\infty} x e^{-x^2} dx \right)
  \left( \int_{\Gamma^{-1/2}}^{\infty} e^{-x^2} dx \right)^{-1}
\label{va1}
\end{equation}

In table \ref{table} we give the numerical values of $f_m$, $f_E$ and $v_a$
computed for three representative values of $\Gamma$.
\begin{table}
\caption{}
\label{table}
\begin{tabular}{lccc}
~~ & $\Gamma=0.3$ & $\Gamma=0.5$ & $\Gamma =1.$\\
$f_m$  & 0.010 & 0.045 & 0.157 \\
$f_E$  & 0.025 & 0.131 & 0.572 \\
$v_a/v_{\rm esc}$ & 1.12  & 1.19 & $1.32$ \\
\end{tabular}
\end{table}
In order to be able to compare these results to observations we
need yet another parameterization. We assume that the rate at which
mass is "processed" in consecutive SPLASHes in the BL, $\dot M_{\rm sp}$
is a fraction $1>\chi>0$ of the mass accretion rate $\dot M_{\rm acc}$,
that is, $\dot M_{\rm sp}=\chi \dot M_{\rm acc}$.
With this we can express the observationally determined or inferred
quantities of outflows (see Introduction) as follows.
For the outflow's mass loss we get
$ \dot M_{\rm o} = \chi f_m \dot M_{\rm acc}$,
its kinetic luminosity $L_{\rm kin} = 0.5 \chi f_E L_{\rm acc}$ and
its momentum discharge $\dot P_{\rm o}= \chi f_m \dot M_{\rm acc} v_a$.
Here $L_{\rm acc} = GM\dot M_{\rm acc}/R$ is the total
accretion luminosity.
We shall discuss the constraints imposed on our parameters by the
observational results in the next section.

We conclude this section by remarking that our derivation here,
although similar to that of Tobertt (1984),
differs in the details of calculation, as explained above, and in
some of the results.

\section{The constraints on the proposed model}
\subsection{General considerations}

The mechanism studied in the previous section transfers
a large fraction of the energy and momentum to a small fraction
of the accreted material.
To reconcile the results of this model with observations we must require that
\begin{itemize}
\item
The mass outflow rate in the wind
$\dot M_{\rm o}= \chi f_m \dot M_{\rm acc}$ is
of the order of 0.01-0.1 (Cabrit 2002 and references therein).
This means that we must have
\begin{equation}
0.01 \la \chi f_m \la 0.1 .
\label{confm}
\end{equation}
\item
The kinetic luminosity of the outflow $L_{\rm kin}= 0.5 \chi f_E L_{\rm acc}$
is $\sim 0.1 L_{\rm bol}$ and $L_{\rm bol} \sim L_{\rm acc}$
(see Introduction and references therein). Thus we must have
\begin{equation}
\chi f_E \sim 0.2 .
\label{confe}
\end{equation}
\item
The momentum discharge of the outflow
$\dot P_{\rm o}=\chi f_m \dot M_{\rm acc} v_a$
is a factor 100 to 1000 larger than the radiation pressure thrust
of the central source $L_{\rm bol}/c$ (see Introduction). With
$L_{\rm bol} \sim L_{\rm acc}= GM \dot M_{\rm acc}/R =
0.5 v_{\rm esc}^2 \dot M_{\rm acc}$
we must have $50 \la \chi f_m v_a c /v_{\rm esc}^2 \la 500$.
Since we have found that $v_a \sim v_{\rm esc}$ for all reasonable values
of $\Gamma$ and since $v_{\rm esc}$ is
typically $\sim 10^{-3} c$ we get $0.05 \la \chi f_m \la 0.5$.
As the velocities entering into the observational determination of
$\dot P_{\rm o}$ are frequently measured rather far from the source,
they are actually lower than our $v_a$ (the mass is larger due to
entrainment). Also, the momentum rate itself can be larger than the initial
momentum if a high pressure
bubble is formed by the shocked fast jet.
Therefore, the last condition on $\chi$ may be actually an overestimate,
and thus the condition from the momentum discharge $\dot P_{\rm o}$ is
in fact similar to the condition from the mass outflow (eq. \ref{confm}).
\end {itemize}

These constraints can be met, for example, by $\Gamma \simeq 0.75$
(giving  $f_m \simeq 0.1$, $f_E \simeq 0.3$, and $v_a \simeq 1.3$)
and $\chi \simeq 0.5$. In this case 95\% of the mass being accreted
does not escape and ultimately has to settle onto the
star, liberating more energy in radiation.
Eventually, most of the energy is liberated via radiation,
as required by observations (e.g., Hartmann 1998).

The main factors which determine whether mass will be ejected,
and if yes what fraction of $\dot M_{\rm acc}$ it is,
are the time scales for the dissipation of the kinetic energy of the
orbiting material in a SPLASH, $\tau_{\rm diss }$, its ejection
(or expansion) time $\tau_{\rm ej}$ and the SPLASH cooling time.
Assuming optically thick conditions in the BL
the cooling time scale is given by the time required for the radiation to
diffuse outward, $\tau_{\rm diff}$.

Note that the above considerations are at all
valid if indeed a SPLASH is formed and, as we have already remarked,
this depends on the expansion and cooling times of individual blobs,
(see eq. \ref{splashform}) out of which the SPLASH is formed. For the sake of
convenience we denote the relevant time scales of an individual blob by $t$
and that of  SPLASH by $\tau$.

The typical expansion (ejection) time of the outflow can be estimated to be
\begin{equation}
\tau_{\rm ej} = H/v_{\rm esc}  \approx \epsilon \Omega_K^{-1},
\label{tauej}
\end{equation}
where the value of the Keplerian angular velocity is calculated for
the BL radius and $\epsilon = H/R$ as defined before.

The dissipation time is more difficult to determine, but we may reasonably
assume that it is of the order of the expansion time of an individual
blob, that is
\begin{equation}
\tau_{\rm diss} \sim t_{\rm ad} = \frac{\ell}{c_s} \approx \frac{\ell}{H}
\Omega_K^{-1},
\label{taudiss}
\end{equation}
where $t_{\rm ad}$ is an individual blob's adiabatic expansion time
and $\ell$ is its size, as discussed before.

The radiative diffusion time depends on the conditions in the disk BL
and can be determined from the photon mean free path $\lambda$ and the
size region from which the photons have to escape and is
\begin{equation}
\tau_{\rm diff} = \epsilon^2 \frac{R^2}{\lambda c},
\label{taudiff}
\end{equation}
where $c$ is the speed of light. Here we have assumed, consistently with our
discussion in the previous section that the size of a SPLASH is of the order
of the BL width, $b$, which, in turn, is of the order of the disk half-thickness
$H=\epsilon R$.

Generally speaking, for the SPLASH mechanism to work we require that in addition
to the condition for the SPLASH to form (see above) we also have
\begin{equation}
\tau_{\rm diff} \gg \tau_{\rm ej}  \ga \tau_{\rm diss}.
\label{cons1}
\end{equation}
If these conditions are met SPLASHes will accelerate mass out of the plane
as a result of shock dissipation of the orbital motion and eject some fraction of
the accreting matter. This can cause the parameter $\Gamma$ to lie
within the limits set by the observational constraints above,
$0.5 \la \Gamma < 1$, as is required by the constraints above.

In the remainder of this section we shall estimate the relevant time scales and check
under what conditions the requirements in the inequality (\ref{cons1})
are met.

\subsection{Time scale estimates}

In the model proposed by Torbett (1984) the BL is assumed to be
arrested abruptly and shocked by a single strong shock. He found
that the cooling time (by diffusion of radiation) exceeds the
expansion time, and concluded that mass can indeed be ejected.
However, by limiting the expansion factor, until when
radiative cooling dominates over adiabatic cooling, he
found that the fraction of mass being lost is only
$f_m \sim 10^{-3}-10^{-2}$, which is below the observed values.
TG, who conducted a numerical calculation based
on the above scenario, found that radiative loses prevent
mass ejection altogether. In this section we shall explain how
our approach, which is based on global energy considerations and
invokes the formations of SPLASHes originating from expanding small blobs,
may circumvent the above limitations.

Before performing the relevant estimates for our model we would like
to mention a few points
which we perceive as difficulties in TG's model and conclusions.
Although not all details of the numerical simulation are given in
the paper, it appeart that
\begin{enumerate}
\item
TG's adiabatic simulations are stopped too early.
As can be seen from their density and velocity plots,
the kinetic energy of the wind is much smaller than that
of the orbiting gas prior to entering the BL.
Allowing more time for converting thermal energy to kinetic one
would have increased the fraction of ejected matter above the
value of $f_m=0.01$ which they find.
It seems that the same argument applies to the simulations
with radiative losses.
\item
In the simulation with radiative losses TG take the cooling to
evolve according to the local diffusion time of the radiation.
Thus in the outer region the cooling time is very short according
to their prescription.
However, radiation from near the midplane should keep the outer
region hotter for a longer time as it diffuses outward,
allowing more efficient acceleration for a longer time.
\item
The density assumed by TG appears to be too low.
They take densities appropriate for a standard viscous
accretion disk with a very high $\alpha$ value ($\alpha\sim 1$)
(Torbett 1984).
However, if the magneto-rotational instability (Balbus \& Hawley 1991)
is responsible for driving the disk turbulence, a value of $\alpha \sim 0.01$
seems to be more appropriate, increasing the density by a
significant factor. In any case $\alpha=1$ is too large by at least an order
of magnitude. In addition, the location of the ejection
should be in the disk BL, in which the physical conditions
may be quite different that the ones in the outer disk.
\end{enumerate}

In our scenario we propose to increase the efficiency of the acceleration
by invoking conditions which increase the radiative
diffusion time, hence reducing the radiative losses.
This can be achieved by considering SPLASHes (as defined section 1.3).
These strongly shocked regions are thus denser than the average disk
density in the BL. The SPLASHes occur in a hot BL and their size,
which we have postulated to be of the order of the BL size, is thus
$\sim H$ (Popham et al. 1993, Regev \& Bertout 1995).

Like Torbett (1984) and TG we assume that the SPLASH cooling time
is the radaitive diffusion time (eq. \ref{taudiff}).
Since $\lambda = (\kappa \rho)^{-1}$, where $\kappa$ is the opacity,
we get
\begin{equation}
\tau_{\rm diff} = \epsilon^2\, R^2\, \frac{\rho \kappa}{c} \simeq
6500 \left( \frac{\epsilon}{0.1} \right)^2
\left(\frac{R} {2 R_{\sun}} \right)^2
 \left(\frac{\rho}{10^{-6} \g \cm^{-3}} \right)
\left(\frac{\kappa}{1 \cm^2 \g^{-1}} \right) \s ,
\label{taudiff2}
\end{equation}
where as before $\epsilon = H/R$.

The scalings in the above equation have been chosen to be appropriate
for a SPLASH in an accretion disk BL around a $\sim 1 M_{\sun}$ star, at
the stellar surface $r\sim 2 R_{\sun}$ (values typical for a TTauri star).
If we take values for a hot BL from models calculated using the method
of Regev \& Bertout (1995), we get for $\alpha=0.05$ and a mass accretion
rate of $\dot M_{\rm acc} = 10^{-7} M_{\sun} \yr^{-1}$ a typical BL density of
$ \sim {\rm ~ 2 \times 10^{-7} \g \cm^{-3}}$ and temperature $\sim 20000 K$.
The opacity in these conditions
is of the order of $100 \cm^2 \g^{-1}$. The value of
$\epsilon$ is  $\sim 0.07$.
See also in Popham (1997), who gives the details of a
similar model of Popham et al. 1993. However
SPLASHes, i.e. strongly shocked regions in the BL, bound to be much hotter
and significantly denser that the BL ambient medium. Therefore choosing
$\rho =10^{-6} \g \cm^{-3}$ and an opacity of the order of the electron
scattering value in (\ref{taudiff2}) seems appropriate.

The expansion time can be conveniently scaled using the escape
velocity version of eq. (\ref{tauej})
\begin{equation}
\tau_{\rm ej} \simeq 500
\left( \frac{\epsilon}{0.1} \right)
 \left(\frac{R}{2 R_{\sun}} \right)
  \left(\frac{v_{\rm esc}}{300 \km \s^{-1}} \right)^{-1} \s.
\label{tauej2}
\end{equation}

We are now ready to compare the various time scales in order to determine if the
SPLASH model is feasible. Using (\ref{tauej}) and (\ref{taudiss}) in the right
inequality in (\ref{cons1}) we obtain the first condition
\begin{equation}
 \frac{\ell}{H} \la \epsilon,
\label{cond1}
\end{equation}
which is very reasonable. It merely states that the size of an individual blob is
significantly smaller than the size of the disk thickness, which is assumed also
to be the size of a SPLASH. With $\epsilon$ typically being $\sim 0.07$ a SPLASH should form
typically from several hundred blobs.

Substituting (\ref{taudiff2}) and (\ref{tauej2}) into the left part of the inequality
(\ref{cons1}) the second condition is obtained

\begin{equation}
14     \left( \frac{\epsilon}{0.1} \right)
\left(\frac{R}{2 R_{\sun}}\right)
\left(\frac{\rho}{10^{-6} \g cm^{-3}} \right)
\left(\frac{\kappa}{1 \cm^2 \g^{-1}} \right)
 \left(\frac{v_{\rm esc}}{300 \km \s^{-1}} \right)
\gg    1  .
\label{cond2}
\end{equation}
With $\epsilon =0.07$ the number in the left hand side of this inequality is
$\approx 10$ and thus we have the strong inequality reasonably satisfied
for the values chosen in the scaling.

It remains to be shown that a SPLASH itself can be formed. For that we need
that the condition expressed in eq. (\ref{splashform}) be fulfilled.
The buildup of a SPLASH starts from small blobs which are
expected to exist in the large sheared BL.
They are weakly shocked to conditions only slightly hotter and denser than those in the BL,
for which the sound speed is $c_s \sim 20 \km \s^{-1}$, and the opacity
is rather large $\kappa \sim 10-1000 \cm^2 \g^{-1}$.
A typical blob size is $\ell \sim \epsilon^2 R$ (from condition \ref{cond1}) and using
this together with a density of $\rho \sim 3 \times 10^{-7} \g \cm^{-3}$ and opacity
of $\kappa \sim 100 \cm^2 \g^{-1}$ in scaling the blob's cooling time by diffusion
of radiation and adiabatic expansion time gives
\begin{equation}
t_{\rm cool} = \frac{\ell^2 \kappa \rho}{c} =
2000  \left( \frac{\epsilon}{0.1} \right)^4
\left(\frac{R} {2 R_{\sun}} \right)^2
 \left(\frac{\rho}{3\, 10^{-7} \g \cm^{-3}} \right)
\left(\frac{\kappa}{100 \cm^2 \g^{-1}} \right) \s ,
\label{tcoolblob}
\end{equation}
and
\begin{equation}
t_{\rm ad} = \frac{\ell}{c_s}= 700
\left( \frac{\epsilon}{0.1} \right)^2
\left(\frac{R} {2 R_{\sun}} \right)
\left(\frac{c_s}{20 \km \s^{-1}}\right)^{-1} \s.
\label{tadblob}
\end{equation}
With $\epsilon =0.07$ the above estimates show that the condition for the formation
of SPLASHES (eq. \ref{splashform}) is satisfied for typical conditions in the disk BL.

It should be stressed again that we have evaluated the conditions in the BL
(and based on them the assumptions about the conditions in a blob and in a splash)
for a particular mass accretion rate - $\dot M_{\rm acc} = 10^{-7} M_{\sun} \yr^{-1}$
and $\alpha$-parameter value $\alpha=0.05$.
Calculations of structure of accretion disk BLs show that the dependence of
the temperature on $\dot M_{\rm acc}$ and $\alpha$ is rather weak.
However, they influence the density $\rho$ approximately as $\propto
\dot M_{\rm acc}^{0.5}$ and $\propto \alpha^{-0.7}$, respectively
(for an opacity law appropriate for YSO disk conditions).
Thus, any mass accretion rate significantly below say $10^{-8} M_{\sun} \yr^{-1}$
or $\alpha$ above say $0.5$ should result in a significantly lower density,
violating condition (\ref{cond2}) and making the thermal acceleration
mechanism not feasible.
We have already pointed out, when criticizing TG's results in the beginning of
this section, that their choice of $\alpha=1$ is in our view
one of the reasons for the failure of their thermal acceleration calculation.

\section {Summary}

This study was motivated by new results from recent X-ray observations
of YSO, which show that there are no differences between the properties
of X-ray emission from YSO with and without outflows (Getman et al. 2002).
The typical X-ray luminosity found by Getman et al.\ (2002)
of systems with outflows is $\sim 2$ orders of magnitude below
the typical kinetic energy of YSO outflows.
We argue that these findings impose severe constraints on any
model in which magnetic fields play dominant dynamical role in
launching the collimated outflows
We have therefore reexamined a previously proposed model for
the acceleration of collimated outflows by
thermal-pressure (Torbett 1984; TG).
Our scenario does not address the collimation process,
for which we accept that magnetic fields have to become dynamically
important, albeit only at large distances from the source.

The main findings of this paper can be summarized as follows.
\begin{enumerate}
\item
 We have performed analytical calculation of the acceleration
of gas from near the boundary layer (BL) of a YSO accretion disk.
The idea is similar, but not identical, to that of Torbett (1984).
By applying global energy conservation we have found that more
mass is able to escape the system than what was found by Torbett (1984).

\item
The fraction of mass escaping the system depends on the
diffusion time of the radiation, and on the fraction $\chi$
of the accreted mass that is strongly shocked in SPLASHes (see below).
For example, if all the dissipated kinetic energy of the
shocked gas goes to the kinetic energy of expansion
in the vertical direction, i.e., $\Gamma=1$,
then a fraction of $f_m=0.16$ of the shocked mass is expelled,
carying with it $f_E = 0.57 $ of the initial energy (see Table 1).
As can be seen in that Table (see section 3.1) if radiative losses
are not severe, the thermal acceleration can account for a
fraction of $\sim 5-10 \%$ of the accreted mass to be blown in a
collimated outflow.
In any case, most of the accreted energy is radiated away,
rather than being in the kinetic energy of this flow.
Magnetic fields are weak at and near the launching region,
but their relative (to thermal pressure and kinetic energy density)
strength becomes appreciable at large distances, where they
probably further collimate the flow.

\item
A crucial ingredient of the proposed model is that the accreted material
will is strongly shocked.
A similar process that can lead to strong shocks in the BL was proposed
(for accretion onto white dwarfs) by Pringle \& Savonije (1979).
Small blobs that are likely to be formed in the strongly sheared BL
are weakly shocked at first. We have shown (eq. \ref{tcoolblob}) that
such blobs cool rather slowly, hence they expand and
collide with more similar orbiting expanding blobs.
Hundreds of such blobs eventually cause a strong disturbance in the BL,
where the accreted matter being strongly shocked, a process we call
SPLASH (spatiotemporally localized accretion shocks).
For the proposed scenario to work the cooling time of the
strongly shocked region should be longer than the ejection time
of matter from this region, and these should be longer than the
dissipation time of the shocked material in the disk (eq. \ref{cons1}).
We have shown (section 3.2) that these conditions are indeed met
in YSO known to blow jets.

\item
The thermal acceleration mechanism works only when the accretion rate
into YSO is $\dot M_{\rm acc} \ga 10^{-8} M_\odot \yr^{-1}$ and
$\alpha \la 0.05$. High $\alpha$ values (which are thought to
be unrealistic on theoretical grounds) prevent this mechanism to work.
To blow jets at the lower range of the allowed accretion rates
($\la 10^{-8} M_\odot \yr$), the
proposed scenario requires SPLASHes in much denser than average density.
This may lead to a prediction of our scenario:
In YSO which blow jets and which have low accretion rate
$\la 10^{-8} M_\odot \yr$ , the ejection
of jets is highly variable on short time scales (of $\sim$ a few hours.
This may be observed as highly variable emission and blobby jet
structure close to the central star.
\end{enumerate}

We thank an anonymous referee whose comments helped to improve this paper.
{}

\end{document}